\documentstyle[twoside,fleqn,espcrc2]{article}

\newcommand{\half}{{\mbox{$\frac{1}{2}$}}}
\newcommand{\be}{\begin{equation}}
\newcommand{\ee}{\end{equation}}
\newcommand{\order}{{\cal O}}

\newcommand{\nr}{{\rm NR}}
\newcommand{\pv}{{\bf p}}

\newcommand{\kin}{{\rm kin}}
\newcommand{\eq}[1]{Eq.~(\ref{#1})}

\newcommand{\Bv}{{\bf B}}
\newcommand{\Ev}{{\bf E}}
\newcommand{\Dv}{{\bf D}}
\newcommand{\delv}{{\bf \del}}
\newcommand{\sigmav}{{\mbox{\boldmath$\sigma$}}}
\newcommand{\msb}{{\overline{\rm MS}}}
\newcommand{\nl}{\nonumber \\}
\newcommand{\delfour}{{\Delta^{(4)}}}
\newcommand{\Mbz}{{M_b^0}}

\newcommand{\Ups}{\Upsilon}

\newcommand{\del}{{\bf \Delta}}
\newcommand{\delsq}{\Delta^{(2)}}

\newcommand{\ie}{{\rm i.e.\ }}

\newcommand{\loc}{{\it l}}
\newcommand{\ainv}{a^{-1}}

\newcommand{\AmS}{{\protect\the\textfont2
  A\kern-.1667em\lower.5ex\hbox{M}\kern-.125emS}}

\title{Precision $\Ups$ Spectroscopy and Fundamental Parameters From NRQCD}

\author{The NRQCD Collaboration
        \thanks{Based on posters presented by G.P. Lepage and J. Sloan.
Members of the collaboration are
C.T.H. Davies and A. Lidsey, Univ. of Glasgow;
K. Hornbostel, SMU;
A. Langnau and G.P. Lepage, Cornell Univ.;
C. Morningstar, Univ. of Edinburgh;
J. Shigemitsu, Ohio State Univ.;
and
J. Sloan, Florida State Univ.}
}

\begin{document}

\begin{abstract}
We present results from a high precision NRQCD simulation of the quenched
$\Ups$ system at $\beta = 6$.  We demonstrate a variety of
important lattice techniques, including the perturbative improvement of
actions, tadpole improvement, and multicorrelated fits for extracting the
spectrum of excited states. We present new determinations of $\alpha_s(M_Z)$
and $M_b$, two fundamental parameters of the Standard Model.
\end{abstract}

\maketitle

\section{Introduction}
In this paper we report new results from an accurate numerical simulation of
the spectrum of the $\Upsilon$~family of mesons. (Results for the
$J/\psi$~family are presented in \cite{others}.) Our simulation uses the NRQCD
\cite{gplbat} action for the quarks, including all   relativistic effects
through $\order(M_b v^4)$ where $M_b$ is the $b$-quark's mass and $v$ its
average velocity.  We demonstrate a variety of important lattice techniques,
including the perturbative improvement of actions, tadpole improvement, and
multicorrelated fits for extracting the spectrum of excited states.
Furthermore, the high statistics and small systematic errors of our results
allow us to extract accurate values for two of the fundamental parameters of
the Standard Model: the mass of the $b$-quark, and the strong coupling
constant $\alpha_\msb(M_Z)$.

The $b$-quarks in $\Upsilon$'s are quite nonrelativistic ($v^2$ is about 0.1).
We exploit this in NRQCD by replacing the Dirac action for the quarks with a
Shr\"odinger action. The Schr\"odinger theory is computationally much easier
to solve because it can be treated as an initial-value problem, rather than
a boundary-value problem. Relativistic effects are systematically introduced,
order-by-order in $v^2$, as corrections to the nonrelativistic
action~\cite{gplbat,corn}. These terms are quite similar in form to the
corrections that remove finite-lattice-spacing errors. We include both types
of correction in our simulations. Consequently the dominant source of
systematic error in our results is the gluon action---we use configurations
produced with the standard Wilson action, and no light-quark vacuum
polarization. (Finite volume errors are negligible since $\Upsilon$'s are
much smaller than ordinary mesons.)

A complication in using improved actions such as ours is that each of the
correction terms has its own coupling constant. These new coupling constants
must be computed somehow. In principle they can be computed using
weak-coupling perturbation theory\cite{gplbat,corn}. Our simulations
demonstrate that this is also the case in practice, provided tadpole-improved
perturbation theory is employed. Indeed (tadpole-improved) tree-level
perturbation theory seems quite sufficient for most aspects of
$\Upsilon$~physics, at least at $\beta=6$; and work has begun on the
first-order corrections\cite{morn}. This perturbative control over the
effective action means that there are really only two parameters in the quark
action, the mass and the charge, just as in the continuum theory. Thus our
simulations are truly calculations from first principles, unlike calculations
based upon a QCD-motivated phenomenological model like the quark potential
model.

In what follows, we first summarize the details of the simulation and fitting
procedures. We then focus upon results for the spectrum, $\ainv$, $M_b$, and
$\alpha_\msb(M_Z)$. Finally we comment upon the implications of our work
concerning improved actions and simulations on coarse grids.

\section{The Simulation}

The NRQCD quark lagrangian we used in our simulations was~\cite{corn}
 \be
 \psi^\dagger \left(1\!-\!\frac{aH_0}{2n}\right)^{n}
 U^\dagger_4
 \left(1\!-\!\frac{aH_0}{2n}\right)^{n}\left(1\!-\!a\delta H\right) \psi,
 \ee
where $n=2$, $H_0$ is the nonrelativistic kinetic energy operator,
 \be
 H_0 = - {\delsq\over2\Mbz},
 \ee
and $\delta H$ is the leading relativistic and finite-lattice-spacing
correction,
 \begin{eqnarray}
\delta H
&=& - \frac{(\delsq)^2}{8(\Mbz)^3}\left(1+\frac{a\Mbz}{2n}\right)
    + \frac{a^2\delfour}{24\Mbz} \nl
& & - \frac{g}{2\Mbz}\,\sigmav\cdot\Bv
            + \frac{ig}{8(\Mbz)^2}\left(\delv\cdot\Ev - \Ev\cdot\delv\right)
\nl & & - \frac{g}{8(\Mbz)^2} \sigmav\cdot(\delv\times\Ev - \Ev\times\delv) .
\label{deltaH}
\end{eqnarray}
Here $\delv$ and $\delsq$ are the simple gauge-covariant lattice derivative
and laplacian, while $\delfour$ is a lattice version of the continuum
operator $\sum D_i^4$. We used the standard cloverleaf operators for the
chromo-electric and magnetic fields, $\Ev$ and~$\Bv$.
The entire action was tadpole improved by dividing every link operator~$U_\mu$
by $u_0 \equiv \langle  \mbox{$\frac{1}{3}$}{\rm Tr}U_{\rm
plaq}\rangle^{1/4}$.  Potential models indicate that corrections
beyond~$\delta H$ contribute only of order 5--10~MeV to $\Upsilon$ energies.

The only parameter in the NRQCD action is the bare mass of the quark, $M_b^0$.
We did a complete simulation for each of three masses: $2/a$, $1.8/a$, and
$1.71/a$. Our analysis indicates that the last of these is the closest to
the real mass (see below). However the statistical analysis for this mass is
not yet complete and so most of the results we quote are for $aM_b^0 = 1.8$.
Except where noted otherwise, the difference has negligible effect.

Note that meson energies in NRQCD are nonrelativistic energies and do not
include the rest mass energy. To obtain the rest mass of the~$\Upsilon$, for
example, we measured the momentum dependence of its energy, fitting it to a
form
 \be
 E_\Upsilon(\pv) = E_\nr(\Upsilon) + \frac{\pv^2}{2M_\kin(\Upsilon)} + \cdots
 \label{kinmass}
 \ee
where the kinetic mass~$M_\kin$ is the physical mass of the meson (see below).

Our quark propagators were computed using an
ensemble of quenched configurations obtained from the Staggered Collaboration
\cite{stag}.  This ensemble contains $105$ $16^3\times 24$ configurations,
generated with an unimproved Wilson action at $\beta=6.0$ and gauge fixed to
Couloumb gauge.  Potential models indicate that the errors due to quenching
and to finite-$a$ errors in the gluon action could be order 40~MeV in
$\Upsilon$~energies, making these the most important systematic errors.

We performed all of our measurements from $8$
different origins separated by $L/2$ in the initial time slice, and treated
the propagators at these origins as statistically independent. At each value
of the bare mass,  our spectrum measurements required a total of $17$ heavy
quark propagators, each with $3$ spin-color components at the source and $6$
at the sink,  using about $17$ GFlops-hour of CPU time, $\ie$ about $800$
inversions per  GFlops-hour.

When measuring meson propagators, we used gauge non-invariant smearing
at both source and sink.  With fast fourier transforms, all source-sink
combinations can be measured at a computational cost of one quark-propagator
inversion per source smearing function \cite{aida}.  We used $1S$, $2S$, $3S$,
$1P$, and $2P$ wavefunctions computed in the quark potential model (with the
Richardson potential) as our smearing functions. In addition we used local
sources for both $S$ and $P$ states. We also created smearing functions for
$S$-states with small, nonzero momenta. For each allowed set of quantum
numbers, all source-sink smearing combinations were measured, resulting in a
$4\times 4$ matrix of correlation functions for $S$ states, $3\times 3$ for
$P$ states, and $2\times 2$ for momentum  states.

\section{Fitting}

We used a variety of fitting strategies to obtain the spectrum from
our meson propagators.  For the spectrum of spin-averaged $S$ and
$P$~states we tried two sorts of multiexponential fit to multiple
propagators. In the first we fit multiple exponentials to the set of
propagators
with smeared sources and a local sink: for example, the ground and first
excited state of the $\Upsilon$ were determined by simultaneously fitting
three exponentials to the $1\loc$ and $2\loc$ propagators, where $1\loc$
indicates $1S$~smearing at the  source and local smearing at the sink, and
$2\loc$ indicates $2S$~smearing at the source and local smearing at the
sink. We discarded results from the highest energy state in such fits, since
this is the most susceptible to biasing due to higher states. These are the
propagators for which our statistics are the best; however, the statistical
quality of the results degrades markedly for the excited states.

The second approach we used for spin-averaged spectra was to fit the full
matrix of propagators formed by taking every combination of nonlocal smearing
function at the source and at the sink: for example, the $\Upsilon$ and its
first two excited states were determined by simultaneously fitting three
exponentials to the 11, 12, 13, 21, 22, 23, 31, 32, and 33 propagators formed
using quark-model wavefunctions for the first three $S$-states as smearing
functions. The statistics were poorer for these smeared-smeared propagators
compared with the smeared-local propagators discussed above; but the
fitting is highly overdetermined, and consequently significantly more robust.
This approach gave the best results for highly excited states; it also avoids
spurious plateaus \cite{splat}.

For splittings between two highly correlated states we found that the best
method is to fit a single exponential to the
jackknifed ratio of the two propagators for the two states.  This approach is
useful for computing spin splittings and the kinetic mass of the~$\Upsilon$.

Statistical uncertainties for the fitting parameters were estimated
by varying the parameters until $\delta\chi^2 = 1$. In several cases we
checked this estimate using a bootstrap analysis. The two estimates agreed
in each case tested, suggesting that the statistical fluctuations in our
propagators are at least approximately gaussian.

Our fitting protocol follows  closely the
techniques discussed in \cite{numrep}, including the use of singular value
decomposition  to allow the inversion of the variance matrix with finite
precision arithmetic.
In general, we had very few problems in obtaining successful fits, even
for the second excited $\Ups$ state.  This is due to the
high statistics in our data set, and to our simultaneous fits using many
smearing functions.  In particular, it was important to include a smearing
function for every state we were attempting to fit.

\section{The Spectrum and $\ainv$}

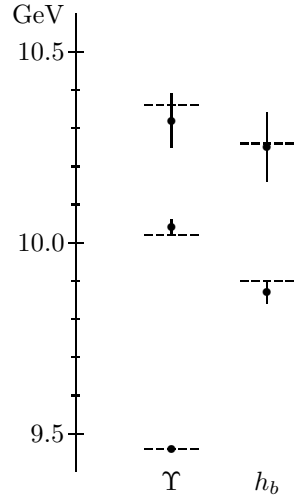
\begin{figure}[t]
\begin{center}
\setlength{\unitlength}{.02in}
\begin{picture}(80,140)(0,930)
% axis
\put(15,940){\line(0,1){120}}
\multiput(13,950)(0,50){3}{\line(1,0){4}}
\multiput(14,950)(0,10){10}{\line(1,0){2}}
\put(12,950){\makebox(0,0)[r]{9.5}}
\put(12,1000){\makebox(0,0)[r]{10.0}}
\put(12,1050){\makebox(0,0)[r]{10.5}}
\put(12,1060){\makebox(0,0)[r]{GeV}}

\put(40,940){\makebox(0,0)[t]{$\Upsilon$}}
\multiput(33,946)(3,0){5}{\line(1,0){2}}
\put(40,946){\circle*{2}}

\multiput(33,1002)(3,0){5}{\line(1,0){2}}
\put(40,1004){\circle*{2}}
\put(40,1004){\line(0,1){2}}
\put(40,1004){\line(0,-1){2}}

\multiput(33,1036)(3,0){5}{\line(1,0){2}}
\put(40,1032){\circle*{2}}
\put(40,1032){\line(0,1){7}}
\put(40,1032){\line(0,-1){7}}

\put(65,940){\makebox(0,0)[t]{$h_b$}}

\multiput(58,990)(3,0){5}{\line(1,0){2}}
\put(65,987){\circle*{2}}
\put(65,987){\line(0,1){3}}
\put(65,987){\line(0,-1){3}}

\multiput(58,1026)(3,0){5}{\line(1,0){2}}
\put(65,1025){\circle*{2}}
\put(65,1025){\line(0,1){9}}
\put(65,1025){\line(0,-1){9}}
\end{picture}
\end{center}
\caption{NRQCD simulation results for the spectrum of the
$\Upsilon (^3S_1)$ and $h_b (^1P_1)$ and their radial excitations. Experimental
values (dashed lines) are indicated for the $S$-states, and for the
spin-average of the $P$-states. The energy zero for the
simulation results is adjusted to give the correct mass to the $\Upsilon$.}
%\caption{Spin averaged $\Ups$ spectrum}
\label{fig:spinav}
\end{figure}

In Figure~\ref{fig:spinav}, we present our simulation results for the
${}^3S_1$ and ${}^1P_1$ spectra, together with the corresponding experimental
values from \cite{pdb}.  We use $\ainv = 2.4$~GeV and Monte Carlo data with
$a\Mbz = 1.8$.  Because the $h_b$ has not been observed, the experimental
values quoted  for the ${}^1P_1$ states are actually the spin averages of the
$\chi_b$ states; strong theoretical arguments, supported by
our measurements, indicate that the two are the same.

\begin{figure}[t]
\begin{center}
\setlength{\unitlength}{0.02in}
\begin{picture}(110,90)(0,-50)
\put(15,-50){\line(0,1){80}}
\multiput(13,-40)(0,20){4}{\line(1,0){4}}
\multiput(14,-40)(0,10){7}{\line(1,0){2}}
\put(12,-40){\makebox(0,0)[r]{$-40$}}
\put(12,-20){\makebox(0,0)[r]{$-20$}}
\put(12,0){\makebox(0,0)[r]{$0$}}
\put(12,20){\makebox(0,0)[r]{$20$}}
\put(12,30){\makebox(0,0)[r]{MeV}}

\multiput(43,-40)(3,0){5}{\line(1,0){2}}
\put(58,-40){\makebox(0,0)[l]{$\chi_{b0}$}}
\put(50,-37){\circle*{2}}
\put(50,-37){\line(0,1){4}}
\put(50,-37){\line(0,-1){4}}

\multiput(23,0)(3,0){5}{\line(1,0){2}}
\put(38,0){\makebox(0,0)[l]{$h_b$}}
\put(30,-4){\circle*{2}}
\put(30,-4){\line(0,1){1}}
\put(30,-4){\line(0,-1){1}}

\multiput(63,-8)(3,0){5}{\line(1,0){2}}
\put(78,-8){\makebox(0,0)[l]{$\chi_{b1}$}}
\put(70,-10){\circle*{2}}
\put(70,-10){\line(0,1){3}}
\put(70,-10){\line(0,-1){3}}

\multiput(83,13)(3,0){5}{\line(1,0){2}}
\put(98,13){\makebox(0,0)[l]{$\chi_{b2}$}}
\put(90,14){\circle*{2}}
\put(90,14){\line(0,1){3}}
\put(90,14){\line(0,-1){3}}
\end{picture}
\end{center}
\caption{Simulation results for the spin structure of the lowest lying
$P$-state in the $\Upsilon$~family. The dashed lines are the experimental
values for the triplet states, and the experimental spin average of all states
for the singlet ($h_b$).}
%\caption{$P$ state hyperfine spectrum}
\label{fig:hyp}
\end{figure}
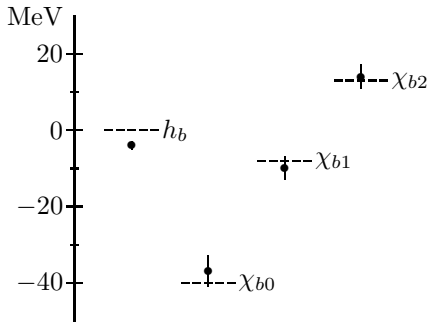

In Figure~\ref{fig:hyp}, we present our results for the spin splittings of
the $P$-wave ground state.  Again $\ainv = 2.4$~GeV and
$a\Mbz = 1.8$. The zero of energy is set to the spin  average of the
$\chi_b(1P)$.     The error bars are only for statistical errors.  We
expect systematic errors in the $P$-state energies of order 5~MeV, which is
comparable to the statistical errors shown.

We determined the inverse lattice spacing by fitting the simulated spectrum
to the experimentally determined spectrum.
Rather than choosing $\ainv$ to make, say,
the $\chi_b(1P) - \Ups(1S)$ splitting correct,
we used a bootstrap analysis to perform a correlated fit for $\ainv$ to
the whole spectrum.  Specifically, treating the $840$ origins in our ensemble
as statistically independent, we generated $20$ bootstrap ensembles, each
containing the meson propagators from $840$ origins (with repetition).
We then extracted the spectrum from each of these ensembles,
giving us an ensemble of $20$~spectra.  Choosing various states for inclusion
in the fit, we did a 1-parameter correlated fit of our spectra to
the experimental spectrum to determine $\ainv$. This procedure was carried out
twice, \ie with two different sets of 20 bootstrap ensembles.

The results of our $\ainv$ fits for different groups of states are summarized
in Table~\ref{tab:ainvfit}. The splittings for $S$-states are between ${}^3S_1$
states, while spin averages are used for the $P$-states. The spin-splittings
used are for the lowest lying $P$-state. For each set of states fit, we
tabulate the values of $\ainv$ and the associated goodness of fit $Q$ for both
sets of $20$~spectra. ($Q$ typically lies between 0.1 and 0.9 for a good fit.)
These fits indicate that
 \be
 \ainv = 2.4(1)~{\rm GeV}.
 \ee
Because of quenching, our $\ainv$ should not agree with those
obtained by matching simulation results for observables that are more infrared,
like the string tension, to experiment; we expect such determinations to yield
smaller values of $\ainv$.
\begin{table}
\begin{center}
\begin{tabular}{lll}
Splittings & $\ainv$ & Q \\ \hline
$2S$-$1S$, $1P$-$1S$ \hspace{3em} & 2.37(9) & .21 \\
& 2.41(10) & .42 \\ \hline
$2S$-$1S$, $1P$-$1S$, & 2.42(7) & .27 \\
$2P$-$1S$ & 2.39(10) & .43 \\ \hline
$2S$-$1S$, $1P$-$1S$, & 2.32(6) & .15 \\
spin-splittings & 2.35(9) & .31 \\ \hline
\end{tabular}
\end{center}
\caption{Bootstrap results for $\ainv$.}
\label{tab:ainvfit}
\end{table}

The success of our global fits of the $\Upsilon$ spectrum show that our
simulation is accurately modeling the general features of the
$\Upsilon$~physics. There are certainly systematic errors in our simulation
results, but these do not degrade the fits because they are generally smaller
than the statistical errors. Only a small improvement in statistics is needed
before these systematic effects will become apparent. Indeed there is already
a slight indication ($2\sigma$) that the $2S$-$1S$ and $1P$-$1S$
splittings are inconsistent, the first being too large and the second too
small. This is precisely the effect expected due to quenching of the gauge
fields. Such discrepancies will be useful when comparable simulations with
unquenched configurations are begun. Similar effects are expected in the spin
splittings, particularly for $S$-states.

\section{$M_b$ determination}
The $b$-quark's mass~$M_b$ is an important fundamental parameter of the
Standard Model. Our simulation results lead to two independent determinations
of this mass (ie, of the pole mass)\cite{mbpap}. These are
among the most accurate and
reliable of all determinations to date.

The first procedure for computing~$M_b$ uses simulation results
for the nonrelativistic energy~$E_\nr(\Upsilon)$ of the $\Upsilon$
(\eq{kinmass}). The quark mass is given by
 \be
 M_b = \half \, \left(M_\Upsilon - \ainv\,(aE_\nr(\Upsilon) - 2aE_0)\right),
 \ee
where $M_\Upsilon = 9.46$~GeV is the experimentally measured mass of
the~$\Upsilon$, and $E_0$ is the nonrelativistic energy of a $\pv\!=\!0$
$b$-quark in NRQCD. The quantity $E_\nr(\Upsilon) - 2E_0$ can be thought of
as the effective binding energy of the meson. The quark energy~$E_0$ is an
ultraviolet divergent quantity and thus it can be computed using
weak-coupling perturbation theory \cite{morn}:
 \be
 aE_0 = b_0\,\alpha_V(q_0)\,\left(1 + \order(\alpha_V)\right) \label{enot}
 \ee
Here  $\alpha_V$ is the strong coupling constant as
defined in \cite{lmac}, and $q_0\sim 1/a$.
In Table~\ref{firstmethod} we present the simulation results for $E_\nr$, and
the corresponding $E_0$'s from perturbation theory.
 \begin{table}
 \begin{center}
 \begin{tabular}{cccc|c}
  $\beta$ & $aM_b^0$ & $aE_\nr(\Upsilon)$ & $aE_0$ & $M_b$\\
 \hline
 6.0  & 1.71 & 0.455(1) &0.21 & 4.69 \\
 & 1.80 & 0.451(1) &0.21 & 4.70 \\
 & 2.00 & 0.444(1)&0.22 & 4.73
 \end{tabular}
 \end{center}
 \caption{Simulation and perturbative results used in the first method for
 determining~$M_b$. Values for $M_b$ are in GeV. }
 \label{firstmethod}
 \end{table}
 From these results we conclude that the pole mass of the $b$-quark is $M_b =
4.7(1)$~GeV. The major source of uncertainty in this determination is from
the two-loop corrections to $E_0$. These could be 10--20\% of $E_0$, or
1--2\% of $M_b$. Uncertainties due to finite lattice spacing, quenching,
tuning of the bare quark mass, statistics, and $\ainv$ are all less than 1\%.

Our second procedure for determining~$M_b$ is to tune the bare quark
mass~$M_b^0$ until the kinetic mass of the~$\Upsilon$, as computed in the
simulation (\eq{kinmass}), agrees with the measured mass of the~$\Upsilon$.
Then the pole mass of the quark is $M_b = Z_m\,M_b^0$ where renormalization
constant~$Z_m$ is computed using perturbation theory \cite{morn}:
 \be
 Z_m = 1 + b_m \alpha_V(q_m) + \order(\alpha_V^2) \label{zm}
 \ee
where scale~$q_m\sim 1/a$. To reduce the sensitivity of the result to the
value of the lattice spacing and to the bare quark mass, we rewrite the
expression for the pole mass as
 \be
 M_b = Z_m\,M_\Upsilon\,\frac{aM_b^0}{aM_\kin(\Upsilon)}
 \ee
where $M_\Upsilon = 9.46$~GeV is the~$\Upsilon$'s experimental mass, while
$M_\kin(\Upsilon)$ is its mass as determined from our simulation with bare
mass~$M_b^0$.
Our results are summarized in Table~\ref{secondmethod}. These indicate that
our best estimate for bare quark mass is
$M_b^0 = 1.7(1)/a$, and therefore the pole mass is again $M_b = 4.7(1)$~GeV.
 \begin{table}
 \begin{center}
  \begin{tabular}{cccc|cc}
  $\beta$ & $aM_b^0$ & $aM_\kin(\Upsilon)$ & $Z_m$ &
  $M_\kin(\Upsilon)$ & $M_b$
  \\ \hline
  6.0 & 1.71 & 3.95(5) & 1.15 &  9.5(4) & 4.72 \\
      & 1.80 & 4.10(5) & 1.15 &  9.8(4) & 4.76 \\
      & 2.00 & 4.45(5) & 1.14 & 10.7(4) & 4.82
  \end{tabular}
 \end{center}
 \caption{Simulation and perturbative results used in the second method for
 determining~$M_b$. Values for $M_\kin(\Upsilon)$ and $M_b$ are in GeV.}
 \label{secondmethod}
 \end{table}
Here the main source of uncertainty is again perturbative: the two-loop
corrections to~$Z_m$ could shift~$M_b$ by 2--3\%. Uncertainties due to
$\ainv$, tuning the bare mass, etc.\ are of order 1\% or less.

Our two determinations of the $b$-quark mass are very different, and yet they
both yield a pole mass of $M_b = 4.7(1)$~GeV.  The complete agreement
between the two methods is a strong indication of the validity of each, and
more generally of the lattice QCD techniques upon which they
rely.

\section{$\alpha_\msb(M_Z)$ Determination}

The strong coupling constant is completely specified by two
lattice quantities:  the inverse
lattice spacing~$\ainv$, and the coupling at a scale $3.41/a$. The inverse
lattice spacing has been discussed above. NRQCD simulations of the $\Upsilon$
system give $\ainv = 2.4(1)$~GeV, where the errors are dominated by statistics.

The coupling constant is obtained by fitting the Monte Carlo value of
$\log\langle \mbox{$\frac{1}{3}$} {\rm Tr}U_{\rm plaq}\rangle$ to the
perturbative expansion\cite{lmac}
  \be
  -\frac{4\pi}{3}\,\alpha_V(3.41/a)\left\{1 - 1.19\alpha_V\right\}.
  \ee
This procedure, in effect, defines $\alpha_V$. The reliability of perturbation
theory can be assessed by computing $\alpha_V$ from a variety of Wilson
loops, each with its own perturbative expansion (corrected to order
$\alpha_V^3$). We have done this and the results are shown in
Table~\ref{loops}. Loops with small areas are less susceptible to
nonperturbative effects and so we conclude that $\alpha_V^{(0)}(3.41/a) =
0.152(1)$.
 \begin{table}
  \begin{center}
  \begin{tabular}{c|c}
  Loop & $\alpha_V(3.41/a)$  \\ \hline
  $\log W_{11} $ & 0.152 \\
  $\log W_{12} $ & 0.152 \\
  $\log W_{13} $ & 0.152 \\
  $\log W_{22} $ & 0.153 \\
  $\log W_{23} $ & 0.154 \\
  $\log W_{33} $ & 0.159
  \end{tabular}
  \end{center}
  \caption{The strong coupling constant as determined from Monte Carlo results
for various Wilson loops.} \label{loops}
 \end{table}

\begin{figure}[t]
 \begin{center}{
% GNUPLOT: LaTeX picture
\setlength{\unitlength}{0.240900pt}
\ifx\plotpoint\undefined\newsavebox{\plotpoint}\fi
\sbox{\plotpoint}{\rule[-0.175pt]{0.350pt}{0.350pt}}%
\begin{picture}(1049,900)(150,0)
\tenrm
\sbox{\plotpoint}{\rule[-0.175pt]{0.350pt}{0.350pt}}%
\put(264,158){\rule[-0.175pt]{173.689pt}{0.350pt}}
\put(264,284){\rule[-0.175pt]{4.818pt}{0.350pt}}
\put(242,284){\makebox(0,0)[r]{$2$}}
\put(965,284){\rule[-0.175pt]{4.818pt}{0.350pt}}
\put(264,410){\rule[-0.175pt]{4.818pt}{0.350pt}}
\put(242,410){\makebox(0,0)[r]{$4$}}
\put(965,410){\rule[-0.175pt]{4.818pt}{0.350pt}}
\put(264,535){\rule[-0.175pt]{4.818pt}{0.350pt}}
\put(242,535){\makebox(0,0)[r]{$6$}}
\put(965,535){\rule[-0.175pt]{4.818pt}{0.350pt}}
\put(264,661){\rule[-0.175pt]{4.818pt}{0.350pt}}
\put(242,661){\makebox(0,0)[r]{$8$}}
\put(965,661){\rule[-0.175pt]{4.818pt}{0.350pt}}
\put(399,158){\rule[-0.175pt]{0.350pt}{4.818pt}}
\put(399,113){\makebox(0,0){$1$}}
\put(399,767){\rule[-0.175pt]{0.350pt}{4.818pt}}
\put(581,158){\rule[-0.175pt]{0.350pt}{4.818pt}}
\put(581,113){\makebox(0,0){$2$}}
\put(581,767){\rule[-0.175pt]{0.350pt}{4.818pt}}
\put(688,158){\rule[-0.175pt]{0.350pt}{4.818pt}}
\put(688,113){\makebox(0,0){$3$}}
\put(688,767){\rule[-0.175pt]{0.350pt}{4.818pt}}
\put(764,158){\rule[-0.175pt]{0.350pt}{4.818pt}}
\put(764,113){\makebox(0,0){$4$}}
\put(764,767){\rule[-0.175pt]{0.350pt}{4.818pt}}
\put(823,158){\rule[-0.175pt]{0.350pt}{4.818pt}}
\put(823,113){\makebox(0,0){$5$}}
\put(823,767){\rule[-0.175pt]{0.350pt}{4.818pt}}
\put(871,158){\rule[-0.175pt]{0.350pt}{4.818pt}}
\put(871,113){\makebox(0,0){$6$}}
\put(871,767){\rule[-0.175pt]{0.350pt}{4.818pt}}
\put(912,158){\rule[-0.175pt]{0.350pt}{4.818pt}}
\put(912,113){\makebox(0,0){$7$}}
\put(912,767){\rule[-0.175pt]{0.350pt}{4.818pt}}
\put(947,158){\rule[-0.175pt]{0.350pt}{4.818pt}}
\put(947,113){\makebox(0,0){$8$}}
\put(947,767){\rule[-0.175pt]{0.350pt}{4.818pt}}
\put(264,158){\rule[-0.175pt]{173.689pt}{0.350pt}}
\put(985,158){\rule[-0.175pt]{0.350pt}{151.526pt}}
\put(264,787){\rule[-0.175pt]{173.689pt}{0.350pt}}
\put(624,23){\makebox(0,0){$q$ (GeV)}}
\put(264,850){\makebox(0,0){$\left(\alpha_V^{(n_f)}(q)\right)^{-1}$}}
\put(764,535){\makebox(0,0)[r]{$n_f=0$}}
\put(764,347){\makebox(0,0)[l]{$n_f=3$}}
\put(340,347){\makebox(0,0){$q^*$}}
\put(264,158){\rule[-0.175pt]{0.350pt}{151.526pt}}
\put(323,234){\usebox{\plotpoint}}
\put(323,234){\usebox{\plotpoint}}
\put(324,235){\usebox{\plotpoint}}
\put(325,236){\usebox{\plotpoint}}
\put(327,237){\usebox{\plotpoint}}
\put(328,238){\usebox{\plotpoint}}
\put(330,239){\usebox{\plotpoint}}
\put(331,240){\usebox{\plotpoint}}
\put(333,241){\usebox{\plotpoint}}
\put(334,242){\usebox{\plotpoint}}
\put(335,243){\usebox{\plotpoint}}
\put(337,244){\usebox{\plotpoint}}
\put(338,245){\usebox{\plotpoint}}
\put(340,246){\usebox{\plotpoint}}
\put(341,247){\usebox{\plotpoint}}
\put(343,248){\usebox{\plotpoint}}
\put(344,249){\usebox{\plotpoint}}
\put(345,250){\usebox{\plotpoint}}
\put(347,251){\usebox{\plotpoint}}
\put(348,252){\usebox{\plotpoint}}
\put(350,253){\usebox{\plotpoint}}
\put(351,254){\usebox{\plotpoint}}
\put(353,255){\usebox{\plotpoint}}
\put(354,256){\usebox{\plotpoint}}
\put(356,257){\rule[-0.175pt]{0.388pt}{0.350pt}}
\put(357,258){\rule[-0.175pt]{0.388pt}{0.350pt}}
\put(359,259){\rule[-0.175pt]{0.388pt}{0.350pt}}
\put(360,260){\rule[-0.175pt]{0.388pt}{0.350pt}}
\put(362,261){\rule[-0.175pt]{0.388pt}{0.350pt}}
\put(364,262){\rule[-0.175pt]{0.388pt}{0.350pt}}
\put(365,263){\rule[-0.175pt]{0.388pt}{0.350pt}}
\put(367,264){\rule[-0.175pt]{0.388pt}{0.350pt}}
\put(368,265){\rule[-0.175pt]{0.388pt}{0.350pt}}
\put(370,266){\rule[-0.175pt]{0.388pt}{0.350pt}}
\put(372,267){\rule[-0.175pt]{0.388pt}{0.350pt}}
\put(373,268){\rule[-0.175pt]{0.388pt}{0.350pt}}
\put(375,269){\rule[-0.175pt]{0.388pt}{0.350pt}}
\put(376,270){\rule[-0.175pt]{0.388pt}{0.350pt}}
\put(378,271){\rule[-0.175pt]{0.388pt}{0.350pt}}
\put(380,272){\rule[-0.175pt]{0.388pt}{0.350pt}}
\put(381,273){\rule[-0.175pt]{0.388pt}{0.350pt}}
\put(383,274){\rule[-0.175pt]{0.388pt}{0.350pt}}
\put(385,275){\rule[-0.175pt]{0.383pt}{0.350pt}}
\put(386,276){\rule[-0.175pt]{0.383pt}{0.350pt}}
\put(388,277){\rule[-0.175pt]{0.383pt}{0.350pt}}
\put(389,278){\rule[-0.175pt]{0.383pt}{0.350pt}}
\put(391,279){\rule[-0.175pt]{0.383pt}{0.350pt}}
\put(392,280){\rule[-0.175pt]{0.383pt}{0.350pt}}
\put(394,281){\rule[-0.175pt]{0.383pt}{0.350pt}}
\put(396,282){\rule[-0.175pt]{0.383pt}{0.350pt}}
\put(397,283){\rule[-0.175pt]{0.383pt}{0.350pt}}
\put(399,284){\rule[-0.175pt]{0.383pt}{0.350pt}}
\put(400,285){\rule[-0.175pt]{0.383pt}{0.350pt}}
\put(402,286){\rule[-0.175pt]{0.383pt}{0.350pt}}
\put(404,287){\rule[-0.175pt]{0.383pt}{0.350pt}}
\put(405,288){\rule[-0.175pt]{0.383pt}{0.350pt}}
\put(407,289){\rule[-0.175pt]{0.383pt}{0.350pt}}
\put(408,290){\rule[-0.175pt]{0.383pt}{0.350pt}}
\put(410,291){\rule[-0.175pt]{0.383pt}{0.350pt}}
\put(411,292){\rule[-0.175pt]{0.396pt}{0.350pt}}
\put(413,293){\rule[-0.175pt]{0.396pt}{0.350pt}}
\put(415,294){\rule[-0.175pt]{0.396pt}{0.350pt}}
\put(416,295){\rule[-0.175pt]{0.396pt}{0.350pt}}
\put(418,296){\rule[-0.175pt]{0.396pt}{0.350pt}}
\put(420,297){\rule[-0.175pt]{0.396pt}{0.350pt}}
\put(421,298){\rule[-0.175pt]{0.396pt}{0.350pt}}
\put(423,299){\rule[-0.175pt]{0.396pt}{0.350pt}}
\put(425,300){\rule[-0.175pt]{0.396pt}{0.350pt}}
\put(426,301){\rule[-0.175pt]{0.396pt}{0.350pt}}
\put(428,302){\rule[-0.175pt]{0.396pt}{0.350pt}}
\put(430,303){\rule[-0.175pt]{0.396pt}{0.350pt}}
\put(431,304){\rule[-0.175pt]{0.396pt}{0.350pt}}
\put(433,305){\rule[-0.175pt]{0.396pt}{0.350pt}}
\put(434,306){\rule[-0.175pt]{0.408pt}{0.350pt}}
\put(436,307){\rule[-0.175pt]{0.408pt}{0.350pt}}
\put(438,308){\rule[-0.175pt]{0.408pt}{0.350pt}}
\put(440,309){\rule[-0.175pt]{0.408pt}{0.350pt}}
\put(441,310){\rule[-0.175pt]{0.408pt}{0.350pt}}
\put(443,311){\rule[-0.175pt]{0.408pt}{0.350pt}}
\put(445,312){\rule[-0.175pt]{0.408pt}{0.350pt}}
\put(446,313){\rule[-0.175pt]{0.408pt}{0.350pt}}
\put(448,314){\rule[-0.175pt]{0.408pt}{0.350pt}}
\put(450,315){\rule[-0.175pt]{0.408pt}{0.350pt}}
\put(451,316){\rule[-0.175pt]{0.408pt}{0.350pt}}
\put(453,317){\rule[-0.175pt]{0.408pt}{0.350pt}}
\put(455,318){\rule[-0.175pt]{0.408pt}{0.350pt}}
\put(457,319){\rule[-0.175pt]{0.460pt}{0.350pt}}
\put(458,320){\rule[-0.175pt]{0.460pt}{0.350pt}}
\put(460,321){\rule[-0.175pt]{0.460pt}{0.350pt}}
\put(462,322){\rule[-0.175pt]{0.460pt}{0.350pt}}
\put(464,323){\rule[-0.175pt]{0.460pt}{0.350pt}}
\put(466,324){\rule[-0.175pt]{0.460pt}{0.350pt}}
\put(468,325){\rule[-0.175pt]{0.460pt}{0.350pt}}
\put(470,326){\rule[-0.175pt]{0.460pt}{0.350pt}}
\put(472,327){\rule[-0.175pt]{0.460pt}{0.350pt}}
\put(474,328){\rule[-0.175pt]{0.460pt}{0.350pt}}
\put(476,329){\rule[-0.175pt]{0.460pt}{0.350pt}}
\put(477,330){\rule[-0.175pt]{0.416pt}{0.350pt}}
\put(479,331){\rule[-0.175pt]{0.416pt}{0.350pt}}
\put(481,332){\rule[-0.175pt]{0.416pt}{0.350pt}}
\put(483,333){\rule[-0.175pt]{0.416pt}{0.350pt}}
\put(484,334){\rule[-0.175pt]{0.416pt}{0.350pt}}
\put(486,335){\rule[-0.175pt]{0.416pt}{0.350pt}}
\put(488,336){\rule[-0.175pt]{0.416pt}{0.350pt}}
\put(490,337){\rule[-0.175pt]{0.416pt}{0.350pt}}
\put(491,338){\rule[-0.175pt]{0.416pt}{0.350pt}}
\put(493,339){\rule[-0.175pt]{0.416pt}{0.350pt}}
\put(495,340){\rule[-0.175pt]{0.416pt}{0.350pt}}
\put(496,341){\rule[-0.175pt]{0.410pt}{0.350pt}}
\put(498,342){\rule[-0.175pt]{0.410pt}{0.350pt}}
\put(500,343){\rule[-0.175pt]{0.410pt}{0.350pt}}
\put(502,344){\rule[-0.175pt]{0.410pt}{0.350pt}}
\put(503,345){\rule[-0.175pt]{0.410pt}{0.350pt}}
\put(505,346){\rule[-0.175pt]{0.410pt}{0.350pt}}
\put(507,347){\rule[-0.175pt]{0.410pt}{0.350pt}}
\put(508,348){\rule[-0.175pt]{0.410pt}{0.350pt}}
\put(510,349){\rule[-0.175pt]{0.410pt}{0.350pt}}
\put(512,350){\rule[-0.175pt]{0.410pt}{0.350pt}}
\put(514,351){\rule[-0.175pt]{0.455pt}{0.350pt}}
\put(515,352){\rule[-0.175pt]{0.455pt}{0.350pt}}
\put(517,353){\rule[-0.175pt]{0.455pt}{0.350pt}}
\put(519,354){\rule[-0.175pt]{0.455pt}{0.350pt}}
\put(521,355){\rule[-0.175pt]{0.455pt}{0.350pt}}
\put(523,356){\rule[-0.175pt]{0.455pt}{0.350pt}}
\put(525,357){\rule[-0.175pt]{0.455pt}{0.350pt}}
\put(527,358){\rule[-0.175pt]{0.455pt}{0.350pt}}
\put(529,359){\rule[-0.175pt]{0.455pt}{0.350pt}}
\put(531,360){\rule[-0.175pt]{0.452pt}{0.350pt}}
\put(532,361){\rule[-0.175pt]{0.452pt}{0.350pt}}
\put(534,362){\rule[-0.175pt]{0.452pt}{0.350pt}}
\put(536,363){\rule[-0.175pt]{0.452pt}{0.350pt}}
\put(538,364){\rule[-0.175pt]{0.452pt}{0.350pt}}
\put(540,365){\rule[-0.175pt]{0.452pt}{0.350pt}}
\put(542,366){\rule[-0.175pt]{0.452pt}{0.350pt}}
\put(544,367){\rule[-0.175pt]{0.452pt}{0.350pt}}
\put(546,368){\rule[-0.175pt]{0.452pt}{0.350pt}}
\put(547,369){\rule[-0.175pt]{0.452pt}{0.350pt}}
\put(549,370){\rule[-0.175pt]{0.452pt}{0.350pt}}
\put(551,371){\rule[-0.175pt]{0.452pt}{0.350pt}}
\put(553,372){\rule[-0.175pt]{0.452pt}{0.350pt}}
\put(555,373){\rule[-0.175pt]{0.452pt}{0.350pt}}
\put(557,374){\rule[-0.175pt]{0.452pt}{0.350pt}}
\put(559,375){\rule[-0.175pt]{0.452pt}{0.350pt}}
\put(561,376){\rule[-0.175pt]{0.482pt}{0.350pt}}
\put(563,377){\rule[-0.175pt]{0.482pt}{0.350pt}}
\put(565,378){\rule[-0.175pt]{0.482pt}{0.350pt}}
\put(567,379){\rule[-0.175pt]{0.482pt}{0.350pt}}
\put(569,380){\rule[-0.175pt]{0.482pt}{0.350pt}}
\put(571,381){\rule[-0.175pt]{0.482pt}{0.350pt}}
\put(573,382){\rule[-0.175pt]{0.482pt}{0.350pt}}
\put(575,383){\rule[-0.175pt]{0.447pt}{0.350pt}}
\put(576,384){\rule[-0.175pt]{0.447pt}{0.350pt}}
\put(578,385){\rule[-0.175pt]{0.447pt}{0.350pt}}
\put(580,386){\rule[-0.175pt]{0.447pt}{0.350pt}}
\put(582,387){\rule[-0.175pt]{0.447pt}{0.350pt}}
\put(584,388){\rule[-0.175pt]{0.447pt}{0.350pt}}
\put(586,389){\rule[-0.175pt]{0.447pt}{0.350pt}}
\put(587,390){\rule[-0.175pt]{0.413pt}{0.350pt}}
\put(589,391){\rule[-0.175pt]{0.413pt}{0.350pt}}
\put(591,392){\rule[-0.175pt]{0.413pt}{0.350pt}}
\put(593,393){\rule[-0.175pt]{0.413pt}{0.350pt}}
\put(594,394){\rule[-0.175pt]{0.413pt}{0.350pt}}
\put(596,395){\rule[-0.175pt]{0.413pt}{0.350pt}}
\put(598,396){\rule[-0.175pt]{0.413pt}{0.350pt}}
\put(600,397){\rule[-0.175pt]{0.482pt}{0.350pt}}
\put(602,398){\rule[-0.175pt]{0.482pt}{0.350pt}}
\put(604,399){\rule[-0.175pt]{0.482pt}{0.350pt}}
\put(606,400){\rule[-0.175pt]{0.482pt}{0.350pt}}
\put(608,401){\rule[-0.175pt]{0.482pt}{0.350pt}}
\put(610,402){\rule[-0.175pt]{0.482pt}{0.350pt}}
\put(612,403){\rule[-0.175pt]{0.482pt}{0.350pt}}
\put(614,404){\rule[-0.175pt]{0.482pt}{0.350pt}}
\put(616,405){\rule[-0.175pt]{0.482pt}{0.350pt}}
\put(618,406){\rule[-0.175pt]{0.482pt}{0.350pt}}
\put(620,407){\rule[-0.175pt]{0.482pt}{0.350pt}}
\put(622,408){\rule[-0.175pt]{0.482pt}{0.350pt}}
\put(624,409){\rule[-0.175pt]{0.442pt}{0.350pt}}
\put(625,410){\rule[-0.175pt]{0.442pt}{0.350pt}}
\put(627,411){\rule[-0.175pt]{0.442pt}{0.350pt}}
\put(629,412){\rule[-0.175pt]{0.442pt}{0.350pt}}
\put(631,413){\rule[-0.175pt]{0.442pt}{0.350pt}}
\put(633,414){\rule[-0.175pt]{0.442pt}{0.350pt}}
\put(634,415){\rule[-0.175pt]{0.482pt}{0.350pt}}
\put(637,416){\rule[-0.175pt]{0.482pt}{0.350pt}}
\put(639,417){\rule[-0.175pt]{0.482pt}{0.350pt}}
\put(641,418){\rule[-0.175pt]{0.482pt}{0.350pt}}
\put(643,419){\rule[-0.175pt]{0.482pt}{0.350pt}}
\put(645,420){\rule[-0.175pt]{0.530pt}{0.350pt}}
\put(647,421){\rule[-0.175pt]{0.530pt}{0.350pt}}
\put(649,422){\rule[-0.175pt]{0.530pt}{0.350pt}}
\put(651,423){\rule[-0.175pt]{0.530pt}{0.350pt}}
\put(653,424){\rule[-0.175pt]{0.530pt}{0.350pt}}
\put(656,425){\rule[-0.175pt]{0.434pt}{0.350pt}}
\put(657,426){\rule[-0.175pt]{0.434pt}{0.350pt}}
\put(659,427){\rule[-0.175pt]{0.434pt}{0.350pt}}
\put(661,428){\rule[-0.175pt]{0.434pt}{0.350pt}}
\put(663,429){\rule[-0.175pt]{0.434pt}{0.350pt}}
\put(664,430){\rule[-0.175pt]{0.482pt}{0.350pt}}
\put(667,431){\rule[-0.175pt]{0.482pt}{0.350pt}}
\put(669,432){\rule[-0.175pt]{0.482pt}{0.350pt}}
\put(671,433){\rule[-0.175pt]{0.482pt}{0.350pt}}
\put(673,434){\rule[-0.175pt]{0.482pt}{0.350pt}}
\put(675,435){\rule[-0.175pt]{0.434pt}{0.350pt}}
\put(676,436){\rule[-0.175pt]{0.434pt}{0.350pt}}
\put(678,437){\rule[-0.175pt]{0.434pt}{0.350pt}}
\put(680,438){\rule[-0.175pt]{0.434pt}{0.350pt}}
\put(682,439){\rule[-0.175pt]{0.434pt}{0.350pt}}
\put(683,440){\rule[-0.175pt]{0.542pt}{0.350pt}}
\put(686,441){\rule[-0.175pt]{0.542pt}{0.350pt}}
\put(688,442){\rule[-0.175pt]{0.542pt}{0.350pt}}
\put(690,443){\rule[-0.175pt]{0.542pt}{0.350pt}}
\put(693,444){\rule[-0.175pt]{0.482pt}{0.350pt}}
\put(695,445){\rule[-0.175pt]{0.482pt}{0.350pt}}
\put(697,446){\rule[-0.175pt]{0.482pt}{0.350pt}}
\put(699,447){\rule[-0.175pt]{0.482pt}{0.350pt}}
\put(701,448){\rule[-0.175pt]{0.385pt}{0.350pt}}
\put(702,449){\rule[-0.175pt]{0.385pt}{0.350pt}}
\put(704,450){\rule[-0.175pt]{0.385pt}{0.350pt}}
\put(705,451){\rule[-0.175pt]{0.385pt}{0.350pt}}
\put(707,452){\rule[-0.175pt]{0.385pt}{0.350pt}}
\put(708,453){\rule[-0.175pt]{0.482pt}{0.350pt}}
\put(711,454){\rule[-0.175pt]{0.482pt}{0.350pt}}
\put(713,455){\rule[-0.175pt]{0.482pt}{0.350pt}}
\put(715,456){\rule[-0.175pt]{0.482pt}{0.350pt}}
\put(717,457){\rule[-0.175pt]{0.642pt}{0.350pt}}
\put(719,458){\rule[-0.175pt]{0.642pt}{0.350pt}}
\put(722,459){\rule[-0.175pt]{0.642pt}{0.350pt}}
\put(725,460){\rule[-0.175pt]{0.482pt}{0.350pt}}
\put(727,461){\rule[-0.175pt]{0.482pt}{0.350pt}}
\put(729,462){\rule[-0.175pt]{0.482pt}{0.350pt}}
\put(731,463){\rule[-0.175pt]{0.482pt}{0.350pt}}
\put(733,464){\rule[-0.175pt]{0.422pt}{0.350pt}}
\put(734,465){\rule[-0.175pt]{0.422pt}{0.350pt}}
\put(736,466){\rule[-0.175pt]{0.422pt}{0.350pt}}
\put(738,467){\rule[-0.175pt]{0.422pt}{0.350pt}}
\put(740,468){\rule[-0.175pt]{0.562pt}{0.350pt}}
\put(742,469){\rule[-0.175pt]{0.562pt}{0.350pt}}
\put(744,470){\rule[-0.175pt]{0.562pt}{0.350pt}}
\put(746,471){\rule[-0.175pt]{0.422pt}{0.350pt}}
\put(748,472){\rule[-0.175pt]{0.422pt}{0.350pt}}
\put(750,473){\rule[-0.175pt]{0.422pt}{0.350pt}}
\put(752,474){\rule[-0.175pt]{0.422pt}{0.350pt}}
\put(754,475){\rule[-0.175pt]{0.562pt}{0.350pt}}
\put(756,476){\rule[-0.175pt]{0.562pt}{0.350pt}}
\put(758,477){\rule[-0.175pt]{0.562pt}{0.350pt}}
\put(760,478){\rule[-0.175pt]{0.482pt}{0.350pt}}
\put(763,479){\rule[-0.175pt]{0.482pt}{0.350pt}}
\put(765,480){\rule[-0.175pt]{0.482pt}{0.350pt}}
\put(767,481){\rule[-0.175pt]{0.422pt}{0.350pt}}
\put(768,482){\rule[-0.175pt]{0.422pt}{0.350pt}}
\put(770,483){\rule[-0.175pt]{0.422pt}{0.350pt}}
\put(772,484){\rule[-0.175pt]{0.422pt}{0.350pt}}
\put(774,485){\rule[-0.175pt]{0.482pt}{0.350pt}}
\put(776,486){\rule[-0.175pt]{0.482pt}{0.350pt}}
\put(778,487){\rule[-0.175pt]{0.482pt}{0.350pt}}
\put(780,488){\rule[-0.175pt]{0.482pt}{0.350pt}}
\put(782,489){\rule[-0.175pt]{0.482pt}{0.350pt}}
\put(784,490){\rule[-0.175pt]{0.482pt}{0.350pt}}
\put(786,491){\rule[-0.175pt]{0.482pt}{0.350pt}}
\put(788,492){\rule[-0.175pt]{0.482pt}{0.350pt}}
\put(790,493){\rule[-0.175pt]{0.482pt}{0.350pt}}
\put(792,494){\rule[-0.175pt]{0.482pt}{0.350pt}}
\put(794,495){\rule[-0.175pt]{0.482pt}{0.350pt}}
\put(796,496){\rule[-0.175pt]{0.482pt}{0.350pt}}
\put(798,497){\rule[-0.175pt]{0.723pt}{0.350pt}}
\put(801,498){\rule[-0.175pt]{0.723pt}{0.350pt}}
\put(804,499){\rule[-0.175pt]{0.402pt}{0.350pt}}
\put(805,500){\rule[-0.175pt]{0.402pt}{0.350pt}}
\put(807,501){\rule[-0.175pt]{0.401pt}{0.350pt}}
\put(809,502){\rule[-0.175pt]{0.482pt}{0.350pt}}
\put(811,503){\rule[-0.175pt]{0.482pt}{0.350pt}}
\put(813,504){\rule[-0.175pt]{0.482pt}{0.350pt}}
\put(815,505){\rule[-0.175pt]{0.602pt}{0.350pt}}
\put(817,506){\rule[-0.175pt]{0.602pt}{0.350pt}}
\put(820,507){\rule[-0.175pt]{0.402pt}{0.350pt}}
\put(821,508){\rule[-0.175pt]{0.402pt}{0.350pt}}
\put(823,509){\rule[-0.175pt]{0.401pt}{0.350pt}}
\put(825,510){\rule[-0.175pt]{0.482pt}{0.350pt}}
\put(827,511){\rule[-0.175pt]{0.482pt}{0.350pt}}
\put(829,512){\rule[-0.175pt]{0.482pt}{0.350pt}}
\put(831,513){\rule[-0.175pt]{0.602pt}{0.350pt}}
\put(833,514){\rule[-0.175pt]{0.602pt}{0.350pt}}
\put(836,515){\rule[-0.175pt]{0.602pt}{0.350pt}}
\put(838,516){\rule[-0.175pt]{0.602pt}{0.350pt}}
\put(841,517){\rule[-0.175pt]{0.402pt}{0.350pt}}
\put(842,518){\rule[-0.175pt]{0.402pt}{0.350pt}}
\put(844,519){\rule[-0.175pt]{0.401pt}{0.350pt}}
\put(846,520){\rule[-0.175pt]{0.482pt}{0.350pt}}
\put(848,521){\rule[-0.175pt]{0.482pt}{0.350pt}}
\put(850,522){\rule[-0.175pt]{0.602pt}{0.350pt}}
\put(852,523){\rule[-0.175pt]{0.602pt}{0.350pt}}
\put(855,524){\rule[-0.175pt]{0.402pt}{0.350pt}}
\put(856,525){\rule[-0.175pt]{0.402pt}{0.350pt}}
\put(858,526){\rule[-0.175pt]{0.401pt}{0.350pt}}
\put(860,527){\rule[-0.175pt]{0.482pt}{0.350pt}}
\put(862,528){\rule[-0.175pt]{0.482pt}{0.350pt}}
\put(864,529){\rule[-0.175pt]{0.602pt}{0.350pt}}
\put(866,530){\rule[-0.175pt]{0.602pt}{0.350pt}}
\put(869,531){\rule[-0.175pt]{0.482pt}{0.350pt}}
\put(871,532){\rule[-0.175pt]{0.482pt}{0.350pt}}
\put(873,533){\rule[-0.175pt]{0.482pt}{0.350pt}}
\put(875,534){\rule[-0.175pt]{0.482pt}{0.350pt}}
\put(877,535){\rule[-0.175pt]{0.602pt}{0.350pt}}
\put(879,536){\rule[-0.175pt]{0.602pt}{0.350pt}}
\put(882,537){\rule[-0.175pt]{0.482pt}{0.350pt}}
\put(884,538){\rule[-0.175pt]{0.482pt}{0.350pt}}
\put(886,539){\rule[-0.175pt]{0.482pt}{0.350pt}}
\put(888,540){\rule[-0.175pt]{0.482pt}{0.350pt}}
\put(890,541){\rule[-0.175pt]{0.482pt}{0.350pt}}
\put(892,542){\rule[-0.175pt]{0.482pt}{0.350pt}}
\put(894,543){\rule[-0.175pt]{0.482pt}{0.350pt}}
\put(896,544){\rule[-0.175pt]{0.482pt}{0.350pt}}
\put(898,545){\rule[-0.175pt]{0.482pt}{0.350pt}}
\put(900,546){\rule[-0.175pt]{0.482pt}{0.350pt}}
\put(902,547){\rule[-0.175pt]{0.482pt}{0.350pt}}
\put(904,548){\rule[-0.175pt]{0.482pt}{0.350pt}}
\put(906,549){\rule[-0.175pt]{0.482pt}{0.350pt}}
\put(908,550){\rule[-0.175pt]{0.482pt}{0.350pt}}
\put(910,551){\rule[-0.175pt]{0.361pt}{0.350pt}}
\put(911,552){\rule[-0.175pt]{0.361pt}{0.350pt}}
\put(913,553){\rule[-0.175pt]{0.964pt}{0.350pt}}
\put(917,554){\rule[-0.175pt]{0.482pt}{0.350pt}}
\put(919,555){\rule[-0.175pt]{0.482pt}{0.350pt}}
\put(921,556){\rule[-0.175pt]{0.361pt}{0.350pt}}
\put(922,557){\rule[-0.175pt]{0.361pt}{0.350pt}}
\put(924,558){\rule[-0.175pt]{0.482pt}{0.350pt}}
\put(926,559){\rule[-0.175pt]{0.482pt}{0.350pt}}
\put(928,560){\rule[-0.175pt]{0.723pt}{0.350pt}}
\put(931,561){\rule[-0.175pt]{0.482pt}{0.350pt}}
\put(933,562){\rule[-0.175pt]{0.482pt}{0.350pt}}
\put(935,563){\rule[-0.175pt]{0.723pt}{0.350pt}}
\put(938,564){\rule[-0.175pt]{0.482pt}{0.350pt}}
\put(940,565){\rule[-0.175pt]{0.482pt}{0.350pt}}
\put(942,566){\rule[-0.175pt]{0.361pt}{0.350pt}}
\put(943,567){\rule[-0.175pt]{0.361pt}{0.350pt}}
\put(945,568){\rule[-0.175pt]{0.723pt}{0.350pt}}
\put(948,569){\rule[-0.175pt]{0.482pt}{0.350pt}}
\put(950,570){\rule[-0.175pt]{0.482pt}{0.350pt}}
\put(952,571){\rule[-0.175pt]{0.723pt}{0.350pt}}
\put(323,234){\usebox{\plotpoint}}
\put(323,234){\rule[-0.175pt]{0.468pt}{0.350pt}}
\put(324,235){\rule[-0.175pt]{0.468pt}{0.350pt}}
\put(326,236){\rule[-0.175pt]{0.468pt}{0.350pt}}
\put(328,237){\rule[-0.175pt]{0.468pt}{0.350pt}}
\put(330,238){\rule[-0.175pt]{0.468pt}{0.350pt}}
\put(332,239){\rule[-0.175pt]{0.468pt}{0.350pt}}
\put(334,240){\rule[-0.175pt]{0.468pt}{0.350pt}}
\put(336,241){\rule[-0.175pt]{0.468pt}{0.350pt}}
\put(338,242){\rule[-0.175pt]{0.468pt}{0.350pt}}
\put(340,243){\rule[-0.175pt]{0.468pt}{0.350pt}}
\put(342,244){\rule[-0.175pt]{0.468pt}{0.350pt}}
\put(344,245){\rule[-0.175pt]{0.468pt}{0.350pt}}
\put(346,246){\rule[-0.175pt]{0.468pt}{0.350pt}}
\put(348,247){\rule[-0.175pt]{0.468pt}{0.350pt}}
\put(350,248){\rule[-0.175pt]{0.468pt}{0.350pt}}
\put(352,249){\rule[-0.175pt]{0.468pt}{0.350pt}}
\put(354,250){\rule[-0.175pt]{0.468pt}{0.350pt}}
\put(355,251){\rule[-0.175pt]{0.466pt}{0.350pt}}
\put(357,252){\rule[-0.175pt]{0.466pt}{0.350pt}}
\put(359,253){\rule[-0.175pt]{0.466pt}{0.350pt}}
\put(361,254){\rule[-0.175pt]{0.466pt}{0.350pt}}
\put(363,255){\rule[-0.175pt]{0.466pt}{0.350pt}}
\put(365,256){\rule[-0.175pt]{0.466pt}{0.350pt}}
\put(367,257){\rule[-0.175pt]{0.466pt}{0.350pt}}
\put(369,258){\rule[-0.175pt]{0.466pt}{0.350pt}}
\put(371,259){\rule[-0.175pt]{0.466pt}{0.350pt}}
\put(373,260){\rule[-0.175pt]{0.466pt}{0.350pt}}
\put(375,261){\rule[-0.175pt]{0.466pt}{0.350pt}}
\put(377,262){\rule[-0.175pt]{0.466pt}{0.350pt}}
\put(379,263){\rule[-0.175pt]{0.466pt}{0.350pt}}
\put(381,264){\rule[-0.175pt]{0.466pt}{0.350pt}}
\put(383,265){\rule[-0.175pt]{0.466pt}{0.350pt}}
\put(384,266){\rule[-0.175pt]{0.500pt}{0.350pt}}
\put(387,267){\rule[-0.175pt]{0.500pt}{0.350pt}}
\put(389,268){\rule[-0.175pt]{0.500pt}{0.350pt}}
\put(391,269){\rule[-0.175pt]{0.500pt}{0.350pt}}
\put(393,270){\rule[-0.175pt]{0.500pt}{0.350pt}}
\put(395,271){\rule[-0.175pt]{0.500pt}{0.350pt}}
\put(397,272){\rule[-0.175pt]{0.500pt}{0.350pt}}
\put(399,273){\rule[-0.175pt]{0.500pt}{0.350pt}}
\put(401,274){\rule[-0.175pt]{0.500pt}{0.350pt}}
\put(403,275){\rule[-0.175pt]{0.500pt}{0.350pt}}
\put(405,276){\rule[-0.175pt]{0.500pt}{0.350pt}}
\put(407,277){\rule[-0.175pt]{0.500pt}{0.350pt}}
\put(409,278){\rule[-0.175pt]{0.500pt}{0.350pt}}
\put(412,279){\rule[-0.175pt]{0.504pt}{0.350pt}}
\put(414,280){\rule[-0.175pt]{0.504pt}{0.350pt}}
\put(416,281){\rule[-0.175pt]{0.504pt}{0.350pt}}
\put(418,282){\rule[-0.175pt]{0.504pt}{0.350pt}}
\put(420,283){\rule[-0.175pt]{0.504pt}{0.350pt}}
\put(422,284){\rule[-0.175pt]{0.504pt}{0.350pt}}
\put(424,285){\rule[-0.175pt]{0.504pt}{0.350pt}}
\put(426,286){\rule[-0.175pt]{0.504pt}{0.350pt}}
\put(428,287){\rule[-0.175pt]{0.504pt}{0.350pt}}
\put(430,288){\rule[-0.175pt]{0.504pt}{0.350pt}}
\put(432,289){\rule[-0.175pt]{0.504pt}{0.350pt}}
\put(435,290){\rule[-0.175pt]{0.530pt}{0.350pt}}
\put(437,291){\rule[-0.175pt]{0.530pt}{0.350pt}}
\put(439,292){\rule[-0.175pt]{0.530pt}{0.350pt}}
\put(441,293){\rule[-0.175pt]{0.530pt}{0.350pt}}
\put(443,294){\rule[-0.175pt]{0.530pt}{0.350pt}}
\put(446,295){\rule[-0.175pt]{0.530pt}{0.350pt}}
\put(448,296){\rule[-0.175pt]{0.530pt}{0.350pt}}
\put(450,297){\rule[-0.175pt]{0.530pt}{0.350pt}}
\put(452,298){\rule[-0.175pt]{0.530pt}{0.350pt}}
\put(454,299){\rule[-0.175pt]{0.530pt}{0.350pt}}
\put(457,300){\rule[-0.175pt]{0.562pt}{0.350pt}}
\put(459,301){\rule[-0.175pt]{0.562pt}{0.350pt}}
\put(461,302){\rule[-0.175pt]{0.562pt}{0.350pt}}
\put(464,303){\rule[-0.175pt]{0.562pt}{0.350pt}}
\put(466,304){\rule[-0.175pt]{0.562pt}{0.350pt}}
\put(468,305){\rule[-0.175pt]{0.562pt}{0.350pt}}
\put(471,306){\rule[-0.175pt]{0.562pt}{0.350pt}}
\put(473,307){\rule[-0.175pt]{0.562pt}{0.350pt}}
\put(475,308){\rule[-0.175pt]{0.562pt}{0.350pt}}
\put(478,309){\rule[-0.175pt]{0.572pt}{0.350pt}}
\put(480,310){\rule[-0.175pt]{0.572pt}{0.350pt}}
\put(482,311){\rule[-0.175pt]{0.572pt}{0.350pt}}
\put(485,312){\rule[-0.175pt]{0.572pt}{0.350pt}}
\put(487,313){\rule[-0.175pt]{0.572pt}{0.350pt}}
\put(489,314){\rule[-0.175pt]{0.572pt}{0.350pt}}
\put(492,315){\rule[-0.175pt]{0.572pt}{0.350pt}}
\put(494,316){\rule[-0.175pt]{0.572pt}{0.350pt}}
\put(497,317){\rule[-0.175pt]{0.512pt}{0.350pt}}
\put(499,318){\rule[-0.175pt]{0.512pt}{0.350pt}}
\put(501,319){\rule[-0.175pt]{0.512pt}{0.350pt}}
\put(503,320){\rule[-0.175pt]{0.512pt}{0.350pt}}
\put(505,321){\rule[-0.175pt]{0.512pt}{0.350pt}}
\put(507,322){\rule[-0.175pt]{0.512pt}{0.350pt}}
\put(509,323){\rule[-0.175pt]{0.512pt}{0.350pt}}
\put(511,324){\rule[-0.175pt]{0.512pt}{0.350pt}}
\put(514,325){\rule[-0.175pt]{0.585pt}{0.350pt}}
\put(516,326){\rule[-0.175pt]{0.585pt}{0.350pt}}
\put(518,327){\rule[-0.175pt]{0.585pt}{0.350pt}}
\put(521,328){\rule[-0.175pt]{0.585pt}{0.350pt}}
\put(523,329){\rule[-0.175pt]{0.585pt}{0.350pt}}
\put(526,330){\rule[-0.175pt]{0.585pt}{0.350pt}}
\put(528,331){\rule[-0.175pt]{0.585pt}{0.350pt}}
\put(531,332){\rule[-0.175pt]{0.516pt}{0.350pt}}
\put(533,333){\rule[-0.175pt]{0.516pt}{0.350pt}}
\put(535,334){\rule[-0.175pt]{0.516pt}{0.350pt}}
\put(537,335){\rule[-0.175pt]{0.516pt}{0.350pt}}
\put(539,336){\rule[-0.175pt]{0.516pt}{0.350pt}}
\put(541,337){\rule[-0.175pt]{0.516pt}{0.350pt}}
\put(543,338){\rule[-0.175pt]{0.516pt}{0.350pt}}
\put(546,339){\rule[-0.175pt]{0.602pt}{0.350pt}}
\put(548,340){\rule[-0.175pt]{0.602pt}{0.350pt}}
\put(551,341){\rule[-0.175pt]{0.602pt}{0.350pt}}
\put(553,342){\rule[-0.175pt]{0.602pt}{0.350pt}}
\put(556,343){\rule[-0.175pt]{0.602pt}{0.350pt}}
\put(558,344){\rule[-0.175pt]{0.602pt}{0.350pt}}
\put(561,345){\rule[-0.175pt]{0.562pt}{0.350pt}}
\put(563,346){\rule[-0.175pt]{0.562pt}{0.350pt}}
\put(565,347){\rule[-0.175pt]{0.562pt}{0.350pt}}
\put(567,348){\rule[-0.175pt]{0.562pt}{0.350pt}}
\put(570,349){\rule[-0.175pt]{0.562pt}{0.350pt}}
\put(572,350){\rule[-0.175pt]{0.562pt}{0.350pt}}
\put(574,351){\rule[-0.175pt]{0.626pt}{0.350pt}}
\put(577,352){\rule[-0.175pt]{0.626pt}{0.350pt}}
\put(580,353){\rule[-0.175pt]{0.626pt}{0.350pt}}
\put(582,354){\rule[-0.175pt]{0.626pt}{0.350pt}}
\put(585,355){\rule[-0.175pt]{0.626pt}{0.350pt}}
\put(587,356){\rule[-0.175pt]{0.482pt}{0.350pt}}
\put(590,357){\rule[-0.175pt]{0.482pt}{0.350pt}}
\put(592,358){\rule[-0.175pt]{0.482pt}{0.350pt}}
\put(594,359){\rule[-0.175pt]{0.482pt}{0.350pt}}
\put(596,360){\rule[-0.175pt]{0.482pt}{0.350pt}}
\put(598,361){\rule[-0.175pt]{0.482pt}{0.350pt}}
\put(600,362){\rule[-0.175pt]{0.578pt}{0.350pt}}
\put(602,363){\rule[-0.175pt]{0.578pt}{0.350pt}}
\put(604,364){\rule[-0.175pt]{0.578pt}{0.350pt}}
\put(607,365){\rule[-0.175pt]{0.578pt}{0.350pt}}
\put(609,366){\rule[-0.175pt]{0.578pt}{0.350pt}}
\put(612,367){\rule[-0.175pt]{0.723pt}{0.350pt}}
\put(615,368){\rule[-0.175pt]{0.723pt}{0.350pt}}
\put(618,369){\rule[-0.175pt]{0.723pt}{0.350pt}}
\put(621,370){\rule[-0.175pt]{0.723pt}{0.350pt}}
\put(624,371){\rule[-0.175pt]{0.530pt}{0.350pt}}
\put(626,372){\rule[-0.175pt]{0.530pt}{0.350pt}}
\put(628,373){\rule[-0.175pt]{0.530pt}{0.350pt}}
\put(630,374){\rule[-0.175pt]{0.530pt}{0.350pt}}
\put(632,375){\rule[-0.175pt]{0.530pt}{0.350pt}}
\put(635,376){\rule[-0.175pt]{0.602pt}{0.350pt}}
\put(637,377){\rule[-0.175pt]{0.602pt}{0.350pt}}
\put(640,378){\rule[-0.175pt]{0.602pt}{0.350pt}}
\put(642,379){\rule[-0.175pt]{0.602pt}{0.350pt}}
\put(645,380){\rule[-0.175pt]{0.530pt}{0.350pt}}
\put(647,381){\rule[-0.175pt]{0.530pt}{0.350pt}}
\put(649,382){\rule[-0.175pt]{0.530pt}{0.350pt}}
\put(651,383){\rule[-0.175pt]{0.530pt}{0.350pt}}
\put(653,384){\rule[-0.175pt]{0.530pt}{0.350pt}}
\put(656,385){\rule[-0.175pt]{0.542pt}{0.350pt}}
\put(658,386){\rule[-0.175pt]{0.542pt}{0.350pt}}
\put(660,387){\rule[-0.175pt]{0.542pt}{0.350pt}}
\put(662,388){\rule[-0.175pt]{0.542pt}{0.350pt}}
\put(665,389){\rule[-0.175pt]{0.803pt}{0.350pt}}
\put(668,390){\rule[-0.175pt]{0.803pt}{0.350pt}}
\put(671,391){\rule[-0.175pt]{0.803pt}{0.350pt}}
\put(674,392){\rule[-0.175pt]{0.542pt}{0.350pt}}
\put(677,393){\rule[-0.175pt]{0.542pt}{0.350pt}}
\put(679,394){\rule[-0.175pt]{0.542pt}{0.350pt}}
\put(681,395){\rule[-0.175pt]{0.542pt}{0.350pt}}
\put(684,396){\rule[-0.175pt]{0.542pt}{0.350pt}}
\put(686,397){\rule[-0.175pt]{0.542pt}{0.350pt}}
\put(688,398){\rule[-0.175pt]{0.542pt}{0.350pt}}
\put(690,399){\rule[-0.175pt]{0.542pt}{0.350pt}}
\put(693,400){\rule[-0.175pt]{0.642pt}{0.350pt}}
\put(695,401){\rule[-0.175pt]{0.642pt}{0.350pt}}
\put(698,402){\rule[-0.175pt]{0.642pt}{0.350pt}}
\put(701,403){\rule[-0.175pt]{0.642pt}{0.350pt}}
\put(703,404){\rule[-0.175pt]{0.642pt}{0.350pt}}
\put(706,405){\rule[-0.175pt]{0.642pt}{0.350pt}}
\put(709,406){\rule[-0.175pt]{0.482pt}{0.350pt}}
\put(711,407){\rule[-0.175pt]{0.482pt}{0.350pt}}
\put(713,408){\rule[-0.175pt]{0.482pt}{0.350pt}}
\put(715,409){\rule[-0.175pt]{0.482pt}{0.350pt}}
\put(717,410){\rule[-0.175pt]{0.642pt}{0.350pt}}
\put(719,411){\rule[-0.175pt]{0.642pt}{0.350pt}}
\put(722,412){\rule[-0.175pt]{0.642pt}{0.350pt}}
\put(725,413){\rule[-0.175pt]{0.642pt}{0.350pt}}
\put(727,414){\rule[-0.175pt]{0.642pt}{0.350pt}}
\put(730,415){\rule[-0.175pt]{0.642pt}{0.350pt}}
\put(733,416){\rule[-0.175pt]{0.562pt}{0.350pt}}
\put(735,417){\rule[-0.175pt]{0.562pt}{0.350pt}}
\put(737,418){\rule[-0.175pt]{0.562pt}{0.350pt}}
\put(739,419){\rule[-0.175pt]{0.562pt}{0.350pt}}
\put(742,420){\rule[-0.175pt]{0.562pt}{0.350pt}}
\put(744,421){\rule[-0.175pt]{0.562pt}{0.350pt}}
\put(746,422){\rule[-0.175pt]{0.843pt}{0.350pt}}
\put(750,423){\rule[-0.175pt]{0.843pt}{0.350pt}}
\put(754,424){\rule[-0.175pt]{0.562pt}{0.350pt}}
\put(756,425){\rule[-0.175pt]{0.562pt}{0.350pt}}
\put(758,426){\rule[-0.175pt]{0.562pt}{0.350pt}}
\put(760,427){\rule[-0.175pt]{0.482pt}{0.350pt}}
\put(763,428){\rule[-0.175pt]{0.482pt}{0.350pt}}
\put(765,429){\rule[-0.175pt]{0.482pt}{0.350pt}}
\put(767,430){\rule[-0.175pt]{0.843pt}{0.350pt}}
\put(770,431){\rule[-0.175pt]{0.843pt}{0.350pt}}
\put(774,432){\rule[-0.175pt]{0.482pt}{0.350pt}}
\put(776,433){\rule[-0.175pt]{0.482pt}{0.350pt}}
\put(778,434){\rule[-0.175pt]{0.482pt}{0.350pt}}
\put(780,435){\rule[-0.175pt]{0.723pt}{0.350pt}}
\put(783,436){\rule[-0.175pt]{0.723pt}{0.350pt}}
\put(786,437){\rule[-0.175pt]{0.482pt}{0.350pt}}
\put(788,438){\rule[-0.175pt]{0.482pt}{0.350pt}}
\put(790,439){\rule[-0.175pt]{0.482pt}{0.350pt}}
\put(792,440){\rule[-0.175pt]{0.723pt}{0.350pt}}
\put(795,441){\rule[-0.175pt]{0.723pt}{0.350pt}}
\put(798,442){\rule[-0.175pt]{0.723pt}{0.350pt}}
\put(801,443){\rule[-0.175pt]{0.723pt}{0.350pt}}
\put(804,444){\rule[-0.175pt]{0.602pt}{0.350pt}}
\put(806,445){\rule[-0.175pt]{0.602pt}{0.350pt}}
\put(809,446){\rule[-0.175pt]{0.723pt}{0.350pt}}
\put(812,447){\rule[-0.175pt]{0.723pt}{0.350pt}}
\put(815,448){\rule[-0.175pt]{0.402pt}{0.350pt}}
\put(816,449){\rule[-0.175pt]{0.402pt}{0.350pt}}
\put(818,450){\rule[-0.175pt]{0.401pt}{0.350pt}}
\put(820,451){\rule[-0.175pt]{0.602pt}{0.350pt}}
\put(822,452){\rule[-0.175pt]{0.602pt}{0.350pt}}
\put(825,453){\rule[-0.175pt]{0.723pt}{0.350pt}}
\put(828,454){\rule[-0.175pt]{0.723pt}{0.350pt}}
\put(831,455){\rule[-0.175pt]{0.602pt}{0.350pt}}
\put(833,456){\rule[-0.175pt]{0.602pt}{0.350pt}}
\put(836,457){\rule[-0.175pt]{0.602pt}{0.350pt}}
\put(838,458){\rule[-0.175pt]{0.602pt}{0.350pt}}
\put(841,459){\rule[-0.175pt]{0.602pt}{0.350pt}}
\put(843,460){\rule[-0.175pt]{0.602pt}{0.350pt}}
\put(846,461){\rule[-0.175pt]{0.964pt}{0.350pt}}
\put(850,462){\rule[-0.175pt]{0.602pt}{0.350pt}}
\put(852,463){\rule[-0.175pt]{0.602pt}{0.350pt}}
\put(855,464){\rule[-0.175pt]{0.602pt}{0.350pt}}
\put(857,465){\rule[-0.175pt]{0.602pt}{0.350pt}}
\put(860,466){\rule[-0.175pt]{0.482pt}{0.350pt}}
\put(862,467){\rule[-0.175pt]{0.482pt}{0.350pt}}
\put(864,468){\rule[-0.175pt]{0.602pt}{0.350pt}}
\put(866,469){\rule[-0.175pt]{0.602pt}{0.350pt}}
\put(869,470){\rule[-0.175pt]{0.964pt}{0.350pt}}
\put(873,471){\rule[-0.175pt]{0.482pt}{0.350pt}}
\put(875,472){\rule[-0.175pt]{0.482pt}{0.350pt}}
\put(877,473){\rule[-0.175pt]{0.602pt}{0.350pt}}
\put(879,474){\rule[-0.175pt]{0.602pt}{0.350pt}}
\put(882,475){\rule[-0.175pt]{0.964pt}{0.350pt}}
\put(886,476){\rule[-0.175pt]{0.482pt}{0.350pt}}
\put(888,477){\rule[-0.175pt]{0.482pt}{0.350pt}}
\put(890,478){\rule[-0.175pt]{0.964pt}{0.350pt}}
\put(894,479){\rule[-0.175pt]{0.482pt}{0.350pt}}
\put(896,480){\rule[-0.175pt]{0.482pt}{0.350pt}}
\put(898,481){\rule[-0.175pt]{0.482pt}{0.350pt}}
\put(900,482){\rule[-0.175pt]{0.482pt}{0.350pt}}
\put(902,483){\rule[-0.175pt]{0.964pt}{0.350pt}}
\put(906,484){\rule[-0.175pt]{0.482pt}{0.350pt}}
\put(908,485){\rule[-0.175pt]{0.482pt}{0.350pt}}
\put(910,486){\rule[-0.175pt]{0.723pt}{0.350pt}}
\put(913,487){\rule[-0.175pt]{0.964pt}{0.350pt}}
\put(917,488){\rule[-0.175pt]{0.482pt}{0.350pt}}
\put(919,489){\rule[-0.175pt]{0.482pt}{0.350pt}}
\put(921,490){\rule[-0.175pt]{0.723pt}{0.350pt}}
\put(924,491){\rule[-0.175pt]{0.482pt}{0.350pt}}
\put(926,492){\rule[-0.175pt]{0.482pt}{0.350pt}}
\put(928,493){\rule[-0.175pt]{0.723pt}{0.350pt}}
\put(931,494){\rule[-0.175pt]{0.964pt}{0.350pt}}
\put(935,495){\rule[-0.175pt]{0.361pt}{0.350pt}}
\put(936,496){\rule[-0.175pt]{0.361pt}{0.350pt}}
\put(938,497){\rule[-0.175pt]{0.964pt}{0.350pt}}
\put(942,498){\rule[-0.175pt]{0.723pt}{0.350pt}}
\put(945,499){\rule[-0.175pt]{0.361pt}{0.350pt}}
\put(946,500){\rule[-0.175pt]{0.361pt}{0.350pt}}
\put(948,501){\rule[-0.175pt]{0.964pt}{0.350pt}}
\put(952,502){\rule[-0.175pt]{0.723pt}{0.350pt}}
\put(955,572){\circle{12}}
\put(927,559){\circle{12}}
\put(888,524){\circle{12}}
\put(869,521){\circle{12}}
\put(852,506){\circle{12}}
\put(654,421){\circle{12}}
\put(624,380){\circle{12}}
\put(410,280){\circle{12}}
\put(377,266){\circle{12}}
\put(955,572){\usebox{\plotpoint}}
\put(945,572){\rule[-0.175pt]{4.818pt}{0.350pt}}
\put(945,573){\rule[-0.175pt]{4.818pt}{0.350pt}}
\put(927,556){\rule[-0.175pt]{0.350pt}{1.204pt}}
\put(917,556){\rule[-0.175pt]{4.818pt}{0.350pt}}
\put(917,561){\rule[-0.175pt]{4.818pt}{0.350pt}}
\put(888,522){\rule[-0.175pt]{0.350pt}{0.964pt}}
\put(878,522){\rule[-0.175pt]{4.818pt}{0.350pt}}
\put(878,526){\rule[-0.175pt]{4.818pt}{0.350pt}}
\put(869,515){\rule[-0.175pt]{0.350pt}{3.132pt}}
\put(859,515){\rule[-0.175pt]{4.818pt}{0.350pt}}
\put(859,528){\rule[-0.175pt]{4.818pt}{0.350pt}}
\put(852,496){\rule[-0.175pt]{0.350pt}{4.818pt}}
\put(842,496){\rule[-0.175pt]{4.818pt}{0.350pt}}
\put(842,516){\rule[-0.175pt]{4.818pt}{0.350pt}}
\put(654,410){\rule[-0.175pt]{0.350pt}{5.059pt}}
\put(644,410){\rule[-0.175pt]{4.818pt}{0.350pt}}
\put(644,431){\rule[-0.175pt]{4.818pt}{0.350pt}}
\put(624,365){\rule[-0.175pt]{0.350pt}{7.468pt}}
\put(614,365){\rule[-0.175pt]{4.818pt}{0.350pt}}
\put(614,396){\rule[-0.175pt]{4.818pt}{0.350pt}}
\put(410,273){\rule[-0.175pt]{0.350pt}{3.373pt}}
\put(400,273){\rule[-0.175pt]{4.818pt}{0.350pt}}
\put(400,287){\rule[-0.175pt]{4.818pt}{0.350pt}}
\put(377,260){\rule[-0.175pt]{0.350pt}{2.650pt}}
\put(367,260){\rule[-0.175pt]{4.818pt}{0.350pt}}
\put(367,271){\rule[-0.175pt]{4.818pt}{0.350pt}}
\end{picture}
}\end{center}
 \caption{$\left(\alpha_V^{(n_f)}(q)\right)^{-1}$ for the $n_f=0$
and $n_f=3$ theories  having  the same $\Upsilon$ physics. The two-loop
perturbative beta function was used to generate the curves. The data points
are values of $\alpha_V^{(0)}(q)$ determined from Monte Carlo measurements of
various small Wilson loops and Creutz ratios of small loops. These show that
two-loop perturbation theory is reliable over the entire range of $q$'s
shown.}
 \label{unquench}
 \end{figure}
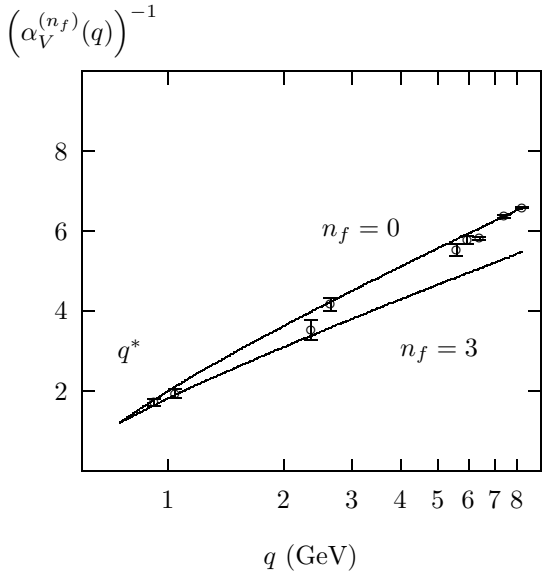

Unfortunately we have $n_f=0$ light-quark flavours in our simulation, while
$\Upsilon$~physics is controlled by a theory with $n_f=3$. Thus a correction
is needed. The correction is illustrated by Figure~\ref{unquench} which shows
plots of the running coupling constants for the $n_f=0$ and $n_f=3$ theories
that give the same $\Upsilon$ physics. An $n_f=0$ simulation gives $n_f=3$
results at scale~$q^*$ if
 \be
 \alpha_V^{(3)}(q^*) = \alpha_V^{(0)}(q^*).
 \ee
For quarkonium the relevant $q^*$ is the mean momentum-transfer
in the potential, which is computed from the expectation value of the
potential between wavefunctions for the state of interest:
 \be
 \langle \frac{\alpha_V^{(3)}(q)}{q^2} \rangle \equiv \alpha_V^{(3)}(q^*)
\langle
 \frac{1}{q^2} \rangle.
 \ee
Quark-model wavefunctions for $1s$, $2s$, and $1p$~states~imply~$q^*$ is
0.75(25)~GeV for $\Upsilon$'s. To correct for the wrong~$n_f$, we use the
two-loop perturbative beta function to make the connections:
 \be
 \begin{array}{ccc}
\alpha_V^{(0)}(3.41/a) && \alpha_V^{(3)}(\mbox{3 GeV}) \\
\downarrow&&\uparrow \\
\qquad\qquad\alpha_V^{(0)}(q^*) &= & \alpha_V^{(3)}(q^*)\qquad\qquad
 \end{array} \nonumber
 \ee
The perturbative beta function is probably reliable down
to scales of order 0.7--0.8~GeV (see Figure~\ref{unquench}). We obtain a
value for the unquenched coupling of
 \be
 \alpha_V^{(3)} (\mbox{3 GeV}) = 0.262(12)
 \ee
where the uncertainty in $q^*$ is the dominant source of error.

It is customary to quote values for the coupling constant in terms of
$\alpha_\msb(M_Z)$. The standard $\msb$~coupling is related to $\alpha_V$ by
 \be
 \alpha_\msb(q) = \alpha_V({\rm e}^{5/6} q) \left\{ 1 + 2 \alpha_V/\pi
  +\order(\alpha_V^2)\right\}.
  \ee
Converting to $\msb$ and running the scale up to the mass of the
$Z$~meson, we obtain finally:
 \be
 \alpha_\msb^{(5)}(M_Z) = 0.112(4),
 \ee
which compares well with other lattice determinations
\cite{aidatalk}, as well as with~0.110--0.125 as obtained from various
high-energy experiments. The main sources of error in our result are:
\begin{center}
 \begin{tabular}{lcc}
$n_f=0 \to 3$ & $\Longrightarrow$ & 0.002 \\
$\alpha_V\to\alpha_\msb$ & $\Longrightarrow$& 0.002 \\
$\alpha_V^{(0)}(3.41/a)$ & $\Longrightarrow$&  0.001 \\
$a^{-1}$  & $\Longrightarrow$ &  0.001
 \end{tabular}
 \end{center}
All of these errors can be significantly reduced in the immediate future.

\section{Perturbative Improvement}

The lowest-order (or Schr\"odinger) NRQCD action accounts for the
bulk of $\Upsilon$ and $\psi$ physics. However high precision results require
an improved action. The key ingredients for an improved action are:
 \begin{itemize}
 \item a tree-level perturbative calculation of the leading correction terms
 in the action (\eq{deltaH});

 \item tadpole improvement of the correction terms, where every link is
replaced by
  \be
  U_\mu \to U_\mu /u_0,
  \ee
and $u_0^4$ is the Monte Carlo value of $\langle \frac{1}{3}{\rm Tr}U_{\rm
plaq}\rangle$; this step is essential for avoiding large renormalizations
of the tree-level corrections;

 \item a calculation of $\order(\alpha_V(1/a))$ renormalizations of the
corrections (if very high precision is needed).
 \end{itemize}
We have implemented the first two of these improvements in our
NRQCD simulations, and they have been strikingly successful, as is illustrated
by the three examples listed in Table~\ref{improved}.
 \begin{table*}
 \begin{center}
 \begin{tabular}{l|ccc|c}
  Quantity & $\delta H = 0$ & $u_0=1$ & Corrected & Exact \\ \hline
 $aM_\Upsilon$ & 3.60(5) & 5.00(5) & 4.15(5) & 4.10(9) \\
 $\chi_{b2} - \chi_{b0}$ & 0 & 25(7) & 51(7) & 53 MeV \\
 $\psi-\eta_c$ & 0 & & 94(1) & 117 MeV \\
 \end{tabular}
 \end{center}
 \caption{NRQCD predictions compared with exact results. Simulation results are
presented for the theory without $\order(a,v)$ corrections ($\delta H=0$),
the theory without tadpole improvement ($u_0=1$), and for the fully corrected
theory.}
 \label{improved}
 \end{table*}

 The first example is the determination of the $\Upsilon$~mass from its
dispersion relation (\eq{kinmass}). This is compared with the ``exact''
result obtained from the relation
 \be
 aM_\Upsilon = 2\left( Z_m\,aM_b^0 - aE_0 \right) + aE_{\rm NR}(\Upsilon),
 \ee
where renormalizations $Z_m$ and $aE_0$ are computed perturbatively, as
discussed above (Eqs.~(\ref{enot}) and~(\ref{zm})). Both the addition of
correction terms to the action and tadpole improvement are essential for
accurate results. The most important correction terms in the action for this
quantity are the spin independent $\order(v^2,a,a^2)$ terms. (Note that the
kinetic mass of a composite particle equals the sum of the masses of the
constituents in a nonrelativistic theory; the binding energy comes  in only
with relativistic corrections.)

The second example is the spin-splitting between the $\chi_{b2}$ and
$\chi_{b0}$ $P$-state in the $\Upsilon$~family. This splitting vanishes in
the lowest order theory since that theory is spin independent. Tadpole
improvement is essential here since the dominant operator that contributes to
this splitting, $\psi^\dagger i\sigmav\cdot \Dv \times g\Ev \psi$, involves the
cloverleaf operator for the chromoelectric field. Again only the fully
corrected theory works well; it gives results in excellent agreement with
experiment.

The third example is the spin-splitting between the~$\psi$ and~$\eta_c$
mesons \cite{others}. The situation here is quite analogous to that for the
$P$-state splittings, except that here the dominant operator is $\psi^\dagger
\sigmav\cdot g\Bv\psi$. Again correction terms and tadpole improvement are
essential. The fully corrected simulation gives an excellent result;  the 20\%
discrepancy between this result and
experiment is consistent with the expected size of $\alpha_V$ corrections,
uncertainties in the mass, $\order(v^4)$ operators, and effects due to
quenching.

These results are compelling evidence that perturbative improvement of
actions, when combined with tadpole-improved perturbation theory, is a highly
effective procedure.

\section{Conclusions}

We have presented early results from our continuing program of
high-precision analyses of heavy-quark mesons using NRQCD.  We have shown
that lattice simulations can accurately account for the
structure of the $\Upsilon$~spectrum up through the $\Upsilon(3S)$, and
including spin structure. And we have used these results to make new
determinations of the strong coupling constant and the $b$-quark's mass.

Our results are possibly the most thorough
QCD tests to date of perturbative improvement for lattice actions. They
underscore the reliability of perturbation theory, the utility of tree-level
improvement, and the critical importance of tadpole improvement (without it,
the improvements tend to be much too small).
The success of the perturbative improvement for NRQCD strongly
suggests that it will work for gluons and light quarks, particularly since the
quarkonium states we examine are 3--5 times smaller than light hadrons. (Note
that $\order(a^2)$ errors for the $\Upsilon$ at $\beta=6$ are only of order a
few percent, even  though the meson's radius is only two lattice spacings.)
Simulations of light hadrons with improved actions should permit precision
work on lattices that are much coarser than those commonly used today.

\section*{Acknowledgements}
This work was supported in part by grants from the DOE, NSF, and SERC.
The computer simulations were performed at the Ohio
Supercomputer Center.


\begin{thebibliography}{9}
\bibitem{others} C.T.H.~Davies, these proceedings; and A.J.~Lidsey, these
  proceedings.
\bibitem{gplbat} G.P. Lepage and B.A. Thacker,
  { Nucl. Phys. B (Proc. Suppl.)} 4 (1988) 199;
  B.A. Thacker and G.P. Lepage, { Phys. Rev.} D43 (1991) 196.
\bibitem{corn}   G.P. Lepage, L. Magnea, C. Nakhleh, U.~Magnea, and
                 K. Hornbostel, { Phys. Rev.} D46 (1992) 4052.
\bibitem{morn}   C. Morningstar, these proceedings.
\bibitem{stag}   R. Gupta {\it et al}, {Phys. Rev.} D43 (1991) 2003.
\bibitem{aida}   A.X. El-Khadra, private communication.
\bibitem{splat}  C.T. Sachrajda, {Nucl. Phys. B (Proc. Suppl.)} 30 (1993) 20.
\bibitem{numrep} W.H.~Press {\it et al}, {\sl Numerical Recipes in C}
 (Cambridge University Press, Cambridge, 1992).
\bibitem{pdb}    Particle Data Group: K. Hikasa {\it et al},
 {Phys. Rev.} D45 (1992).
\bibitem{lmac} G.P. Lepage and P.B. Mackenzie,  {Phys. Rev.} D48 (1993) 2250.
\bibitem{aidatalk} A.X. El-Khadra, these proceedings.
\bibitem{mbpap} NRQCD Collaboration: C.T.H.~Davies {\it et al}, submitted to
Phys. Rev. Lett.
\end{thebibliography}
\end{document}